\documentclass[12pt]{elsarticle}
\usepackage{alltt}
\usepackage{url}
\usepackage{hyperref}

\makeatletter
\renewcommand{\@oddfoot}{\hfill\@date}
\makeatother

\title{\sf\Large An 8- and 12-bit block AES cipher}
\author{\sf Peter T. Breuer}
\date{23 April 2013}

\begin{document}
\sf

\thispagestyle{empty}
\maketitle

\section*{Abstract}
\noindent
{\em
Because it is so unusual,  or hard to find, or expository,
a truly tiny 8- or 12-bit block AES (Rijndael) cipher is documented
here, along with Java source code.
}

\section{Introduction}

The intent here is to document a simple 8- or 12-bit (or more) block AES
(Rijndael) cipher \cite{AES}. The small numbers make for a
very clear exposition.
AES ciphers are standardly of much larger block-size than 8 or 12
bits, with 128 bits or more being typical, but small
block-sizes can still be useful in combination with
appropriate padding of the plaintext and frequent changes of key.

My own difficulty in producing a small AES cipher for use in
encrypted computing \cite{BB13} lay with the unfortunate circumstance that
I had no cryptographic knowledge at all and could not find any
implementation code for the right block-size on the Web.  So I
reverse-engineered those AES source codes that I could find, generating
an understanding of the principles in the process -- 
well enough to create my own tiny version of the cipher.

The source code that helped me most (other codes I found were too
sophisticated) came in a set of files called
'simple\_aes.c', `ecb\_decrypt.c', `simple\_aes\_decr.c',
`ecb\_encrypt.c', without any indicator of authorship,
but I believe from associated hints and a memory of my original search
that the files came from the Cambridge University Computer Lab.  There
is certainly a great deal of expertise evident in that source code --
things are done clearly and simply, which is a hard trick to pull off
if one is not an expert.

The 12-bit block cipher will be described first below, out of no
particular prejudice, and then the few changes that the
8-bit block cipher requires will be described.  A Java source code
implementation for the 8-bit block cipher will be set out too (and an
Appendix to the text here has it verbatim).

\section{The 12-bit cipher}

The idea of the AES cipher is simple -- do a sequence of 1-1 operations
in an appropriate space, the parameters of each operation in the
sequence being set by the secret key.  To decrypt, run the same
sequence of 1-1 operations in reverse.

The `appropriate space' here is a space of 4-dimensional (`4D') vectors.
Each of the four ordinates are numbers in the range 0-7,
i.e.  a 3-bit binary number.  That means that the cipher works with
4$\times$3=12-bit plaintext, converting it to a 12-bit ciphertext. That
is a 12-bit block cipher.
Later we will also look at a 4$\times$2=8-bit design.

Each 3-bit ordinate lives in a special arithmetic field, a so-called
GF(8) field of numbers 0-7 which when written out as three binary bits
$a_2a_1a_0$ can be taken as the coefficients of a polymomial $a_2x^2
+a_1x+a_0$ `over' the field of modulo 2 arithmetic, which means that
the coefficients are mod 2, and are added and multiplied mod 2 when
required.  
Polynomials are added and multiplied in the usual way polynomials are
multiplied, subject to (i) the coefficients are mod 2, and (ii) $x$ is
a root of a certain irreducible cubic polynomial $p(x)$, so $p(x)=0$.
A typical 3-bit ordinate is 101, or $x^2+1$, being 1 times $x^2$ plus 0
times $x^1$ plus 1 times $x^0$.

I choose to set $p(x) = x^3+x+1$, so that $x^3+x+1=0$.
The equation allows powers of 3 and higher of $x$ to be expressed
in terms of the powers of 2 and lower.  That is $x^3 = x+1$, since
the coefficients are taken modulo 2.

One can be sure that my choice of $p(x)$ is irreducible over this field,
because if it were to factor at least one of the factors would be linear
(since the order of $p$ is 3, so the factors must have order 2 and 1, or
1 and 1 and 1).  A linear factor is of the form $x-K$, where $K$ is in
the underlying modulo 2 arithmetic field, hence 0 or 1.  So 0 or 1 would
be a root of $p$.  But $p(0)=1$ and $p(1)=1+1+1=1$, so neither is a
root.  Hence $p$ is indeed irreducible.

In this setting, 3+2=1
because that is 011 and 010 added bitwise to
give 001 mod 2.  And 3*2, i.e., 011*010, is calculated by multiplying
$0x^2+x+1$ by $0x^2+x+0$, giving $x^2+x$, or 110, i.e.  6, as
expected.  But 3*4, i.e., 011*100, is calculated multiplying $0x^2+x+1$
by $x^2$, giving $x^3+x^2$, or $x+1+x^2$, i.e., 111.  Thus 3*4=7.

The reason for choosing GF(8) rather than, say, simple addition and
multiplication modulo 8 on the numbers 0-7, is that multiplication mod 8
has non-zero divisors of zero, so that 4*2=0, for example.  So there
cannot be a division operation mod 8.  In contrast, GF(8) is a field and
every non-zero number has a multiplicative inverse in it.  Being a
field, one can make a matrix with entries in the field, and hope to be
able to invert the matrix provided its determinant is non-zero.

Nevertheless, there might be viable alternatives to a GF($2^n$) field in
the construction here, particularly when $n=2^m$.  I am thinking of the
power-$2^{2^m}$ fragments of Conway's On$_2$ field \cite{JHWC}.

This is the sequence of 1-1 operations that the cipher performs on the
4D vector $\bar{x}$ with 3-bit ordinates in the range 0-7:

\begin{enumerate}
\item add a constant 4D vector $\bar{K}_1$:
$
\bar{x} \mapsto \bar{x}+\bar{K}_1
$
\item apply a constant 1-1 permutation $S$ of 0-7 in-place on each ordinate:
\[
{(x_1,x_2,x_3,x_4)} \mapsto (S(x_1),S(x_2),S(x_3),S(x_4))
\]
\item apply a constant permutation $\pi$ to the ordinates:
\[
{(x_1,x_2,x_3,x_4)} \mapsto (x_{\pi(1)},x_{\pi(2)},x_{\pi(3)},x_{\pi(4)})
\]
\item apply a constant invertible 4D matrix transformation $M$:
$
\bar{x} \mapsto M\bar{x}
$
\item add a constant 4D vector $\bar{K}_2$:
$
\bar{x} \mapsto \bar{x}+\bar{K}_2
$
\item apply the constant 1-1 permutation $S$ of 0-7 in-place on each ordinate
again:
\[
{(x_1,x_2,x_3,x_4)} \mapsto (S(x_1),S(x_2),S(x_3),S(x_4))
\]
\item apply the constant permutation $\pi$ to the ordinates again:
\[
{(x_1,x_2,x_3,x_4)} \mapsto (x_{\pi(1)},x_{\pi(2)},x_{\pi(3)},x_{\pi(4)})
\]
\item add a constant 4D vector $\bar{K}_3$:
$
\bar{x} \mapsto \bar{x}+\bar{K}_3
$
\end{enumerate}
The decryption operation does these operations in reverse, thus:
\begin{enumerate}
\item subtract the constant 4D vector $\bar{K}_3$:
$
\bar{x} \mapsto \bar{x}-\bar{K}_3
$
\item apply the constant permutation $\pi^{-1}$ to the ordinates :
\[
{(x_1,x_2,x_3,x_4)} \mapsto
(x_{\pi^{-1}(1)},x_{\pi^{-1}(2)},x_{\pi^{-1}(3)},x_{\pi^{-1}(4)})
\]
\item apply the constant 1-1 permutation $S^{-1}$ in-place on each
ordinate:
\[
{(x_1,x_2,x_3,x_4)} \mapsto (S^{-1}(x_1),S^{-1}(x_2),S^{-1}(x_3),S^{-1}(x_4)
\]
\item subtract the constant 4D vector $\bar{K}_2$:
$
\bar{x} \mapsto \bar{x}-\bar{K}_2
$
\item apply the constant 4D matrix transformation $M^{-1}$:
$
\bar{x} \mapsto M^{-1}\bar{x}
$
\item apply a constant permutation $\pi^{-1}$ to the ordinates:
\[
{(x_1,x_2,x_3,x_4)} \mapsto
(x_{\pi^{-1}(1)},x_{\pi^{-1}(2)},x_{\pi^{-1}(3)},x_{\pi^{-1}(4)})
\]
\item apply the constant 1-1 substitution $S^{-1}$ in-place on each
ordinate:
\[
{(x_1,x_2,x_3,x_4)} \mapsto (S^{-1}(x_1),S^{-1}(x_2),S^{-1}(x_3),S^{-1}(x_4))
\]
\item subtract the constant 4D vector $\bar{K}_1$:
$
\bar{x} \mapsto \bar{x}-\bar{K}_1
$
\end{enumerate}
With this design, encryption and decryption are guaranteed to be
mutually inverse operations. One may note that the transformation is 
really $A_1;S^4;A_2;S^4;A_3$ where $A_1$, $A_2$, $A_3$ are affine
transformations on the 4D vector space, and $S^4$ is the permutation $S$ of 
the field elements applied in place to each of the four ordinates.

The vectors $\bar{K}_1$, $\bar{K}_2$, $\bar{K}_3$, are derived from a
single 12-bit input key. They are called `round keys', and change as the
input key is changed.  The permutations $S$, $\pi$ and the matrix $M$ are
constant, selected beforehand and coded statically into the algorithm.

For the permutation $S$ of 0-7, I chose one that factors into two
cycles of length 3 and 5. That implies that it
takes 15 applications to repeat itself when applied simultaneously to every
number in the range 0-7. I simply divided the set $\{0\dots 7\}$
into two disjoint sets of size 3 and 5 respectively, and wrote down
an arbitrary cycle on each:
\[
S:\begin{array}[b]{cc}
\{0,2,6\} & \{1,3,4,5,7\}\\[1ex]
0\mapsto 2\mapsto 6\mapsto 0&
1\mapsto 4\mapsto 7\mapsto 5\mapsto 3\mapsto 1
\end{array}
\]
It seemed best to aim at a long orbit for this permutation.

For the permutation $\pi$ of the ordinates 1-4, I simply chose the
operation that swaps ordinates 1 and 3, that is
\[
\pi :~~1\mapsto 3\mapsto 1\quad 2\mapsto 2\quad 4\mapsto 4
\]
For the matrix $M$, I chose 
\[
M = \left(\begin{array}{cccc}
           1&4&0&0\\
           4&1&0&0\\
           0&0&1&4\\
           0&0&4&1
           \end{array}\right)
\]
This form reduces the problem of finding the inverse to
that of inverting only one 2$\times$2 matrix. The formula for the
inverse of a 2$\times$2 matrix 
\[
\left(\begin{array}{cc}
           a&b\\
           c&d
           \end{array}\right)^{-1}
=
\left(\begin{array}{cc}
           d&-b\\
           -c&a
           \end{array}\right)/(ad-bc)
\]
tells us that the inverse is
\[
\left(\begin{array}{cc}
           1&4\\
           4&1
           \end{array}\right)^{-1}
=
\left(\begin{array}{cc}
           1&-4\\
           -4&1
           \end{array}\right)/(1*1-4*4)
=
\left(\begin{array}{cc}
           1&4\\
           4&1
           \end{array}\right)/7
\]
The arithmetic needs to be done in the field GF(8):  in particular,
the determinant $1*1-4*4$ is $1*1-(x^2)^2 = x^4 + 1 = x(x^3)+1 = x(x+1)+1 =
x^2+x+1 $, or 111, or '7'. So we need
to divide by 7 in GF(8). Also $-4=4$ in GF(8). Addition is just mod 2
arithmetic on the individual bits of the three-bit representation. So
$100+100=0$, i.e., $4+4=0$ and hence $-4=4$.

In fact the inverse of 7 in GF(8) is 4, as one may check: $(x^2+x+1)x^2 =
x^4+x^3+x^2 = xx^3+x^3+x^2 = x(x+1)+x+1+x^2 = x^2+x+x+1+x^2 = 1$. So
instead of dividing by 7 one may multiply by 4, and obtain as the
2$\times$2 matrix inverse
\[
\left(\begin{array}{cc}
           4&6\\
           6&4
           \end{array}\right)
\]
since $4*4=6$ in GF(8), being $x^2x^2 = x^4 = x(x^3) = x(x+1) = x^2+x$.
Thus
\[
M^{-1} =
\left(\begin{array}{cccc}
           4&6&0&0\\
           6&4&0&0\\
           0&0&4&6\\
           0&0&6&4
           \end{array}\right)
\]
How do we get the constant 4D vectors $\bar{K}_1$, $\bar{K}_2$, $\bar{K}_3$?
I simply borrowed verbatim a mixing procedure I found in the code I
reverse-engineered.  These so-called `round' keys are generated from
the single secret input 12-bit (i.e., 4D) key $\bar{k}$.

Let the key be $\bar{k} = (k_1,k_2,k_3,k_4)$ as a 4D vector of
ordinates in the range 0-7 from the field GF(8).
Then we make a sequence of 6-bit (i.e. 2D) vectors as follows:
\[
\begin{array}{lcl}
w_0 &= & (k_1,k_2)\\
w_1 &= & (k_3,k_4) = (w_{11},w_{12})\\
w_2 &= & w_0 + (4,0) + (S(w_{12}),S(w_{11})) \\
w_3 &= & w_1 + w_2 = (w_{31},w_{32})\\
w_4 &= & w_2 + (6,0) + (S(w_{32}),S(w_{31})) \\
w_5 &= & w_3 + w_4
\end{array}
\]
That is:
\begin{enumerate}
\item take the top 6 bits of $\bar{k}$ and call it  $w_0$;
\item take the bottom 6 bits of $\bar{k}$ and call it $w_1$;
\item swap the ordinates of $w_1$ and apply $S$ to each, then add
$(4,0)$ and $w_0$, giving $w_2$;
\item add $w_1$ and $w_2$ together, giving $w_3$;
\item swap the ordinates of $w_3$ and apply $S$ to each, then add
$(6,0)$ and $w_2$, giving $w_4$;
\item add $w_3$ and $w_4$ together, giving $w_5$.
\end{enumerate}
Then the three 12-bit `round keys' are
\[
\begin{array}{lcl}
\bar{K}_1 &=& \bar{k}\\
\bar{K_2} &=& (w_2,w_3)\\
\bar{K_3} &=& (w_4,w_5)
\end{array}
\]
I have no particular insight into this procedure. It seems to me that
one might as well use a pseudo-random generator if one has a
computer available, and the procedure above is likely most useful
if one is restricted to hardware implementations.

\section{The 8-bit cipher}

For the 8-bit cipher I use exactly the same setup as described above,
with the difference that the 4D vectors have ordinates in the range
0-3, not 0-7, from the field GF(4), not GF(8). That makes up
4$\times$2=8 bits in total.

I generate the GF(4) field by choosing as irreducible polynomial
$p(x)=x^2+x+1$. That generates the equivalence
$x^2=x+1$ over the  modulo 2 arithmetic field.
Write numbers in the range 0-3 as linear
polynomials over the field mod 2, interpreting the bits of the binary
expansion of the number as the
coefficients of the powers of $x$. Then
addition and multiplication in GF(4) are
addition and multiplication of polynomials, as before.

I chose the permutation $S$ on 0-3 by dividing the set $\{0\dots 3\}$
into two parts of size 3 and 1 respectively, and factoring $S$ as
cycles over the two parts, thus:
\[
S:\begin{array}[b]{cc}
\{0,2,3\} & \{1\}\\[1ex]
0\mapsto 3\mapsto 2\mapsto 0&
1\mapsto 1
\end{array}
\]
This only has an orbit length 3, but I felt it preferable to length 4
or 2, which are the only other options (NB: I later learned that one
ought to avoid $S$ having fixed points. so 2 or 4 -- two
transpositions or one 4-cycle -- would have been better).

The permutation $\pi$ of the four ordinates of the vector
swaps $x_2$ and $x_4$ in a 4D vector $(x_1,x_2,x_3,x_4)$, as before.

The matrix transformation that I chose to use is
\[
\begin{array}{l@{~}c@{~}l@{\qquad}l@{~}c@{~}l}
M&=&\left(\begin{array}{cccc}
     1&2&0&0\\
     2&1&0&0\\
     0&0&1&2\\
     0&0&2&1
    \end{array}\right)&
M^{-1}&=&\left(\begin{array}{cccc}
     3&1&0&0\\
     1&3&0&0\\
     0&0&3&1\\
     0&0&1&3
    \end{array}\right)
\end{array}
\]
Round key generation follows the same pattern as before, but this time
the 2D vectors are 4-bit, since the ordinates are 2-bit, so I had to add
in different constant vectors when generating $w_2$ and $w_4$:
\[
\begin{array}{lcl}
w_0 &= & (k_1,k_2)\\
w_1 &= & (k_3,k_4) = (w_{11},w_{12})\\
w_2 &= & w_0 + (2,0) + (S(w_{12}),S(w_{11})) \\
w_3 &= & w_1 + w_2 = (w_{31},w_{32})\\
w_4 &= & w_2 + (3,0) + (S(w_{32}),S(w_{31})) \\
w_5 &= & w_3 + w_4
\end{array}
\]
and then the round keys $\bar{K}_1$, $\bar{K}_2$, $\bar{K}_3$
are generated by the same procedure as before.

\section{Java source code}

The 8-bit AES code is documented here. The 12-bit code
differs from it only in the ways documented above.

To encrypt a block (which is a single byte long here) we use the
`ecb\_encrypt' method:

\newenvironment{program}{
\begin{footnotesize}\begin{quote}\begin{alltt}
}{
\end{alltt}\end{quote}\end{footnotesize}
}

\begin{program}
byte ecb_encrypt(byte input_block) \{
  state = input_block & 0xff;
  encrypt();
  return (byte)state;
\}
\end{program}

\noindent
The `state' is the working area that is transformed by the cipher.
For convenience it is declared integer, but ony the low 8 bits are ever
used for the 8-bit cipher.  The `encrypt' method transforms the state
according to the cipher.  The state variable really could be declared `byte'
rather than `int' (as it is here), but it is awkward in Java to work
with integers of size other than `int' because compound expressions
are converted up to `int' automatically, so almost every assignment to the
state variable would require a cast back to `byte' again.

Decryption of an 8-bit encrypted block is similar in code form. The
code sets the state variable, then
fires off the `decrypt' method, which changes the state, and the
deciphered state is returned:

\begin{program}
byte ecb_decrypt(byte cipherblock) \{
  state = cipherblock & 0xff;
  decrypt();
  return (byte)state;
\}
\end{program}

\noindent
The `encrypt' method performs the eight steps described in the last
section on the low 8 bits of the `state' integer. Note that the
exclusive-or \verb$^$ operation does bitwise mod 2 addition, which is
the addition in the field, and consequently on the 4-D vectors written
out as sequences of 4 consecutive 2-bit sequences, one for each of their
ordinates:

\begin{program}
void encrypt() \{
  state ^= roundkeys[0];   // 1. add K1
  substitute_nibble();     // 2. recode 2 bit ordinates
  swaprow();               // 3. swap 1,3 ordinates
  mixcolumns();            // 4. multiply by matrix M
  state ^= roundkeys[1];   // 5. add K2
  substitute_nibble();     // 6. recode 2 bit ordinates
  swaprow();               // 7. swap 1,3 ordinates again
  state ^= roundkeys[2];   // 8. add K3
\}  
\end{program}

\noindent
The `decrypt' method is inverse:

\begin{program}
public void decrypt() \{
  state ^= roundkeys[2];   // 1. add K3
  swaprow();               // 2. swap 1,3 ordinates
  inv_substitute_nibble(); // 3. recode 2 bit ordinates
  state ^= roundkeys[1];   // 4. add K2
  inv_mixcolumns();        // 5. multiply by inverse of matrix M
  swaprow();               // 6. swap 1,3 ordinates
  inv_substitute_nibble(); // 7. recode 2 bit ordinates
  state ^= roundkeys[0];   // 8. add K1
\}
\end{program}

\noindent
Adding a 4D vector is the same as subtracting it, since the addition
is bitwise at bottom, and that is modulo 2.

The `swaprows' method interchanges $x_2$ and $x_4$ in the vector
$(x_1,x_2,x_3,x_4)$. It is the permutation $\pi$ applied to
the `state' vector:

\begin{program}
void swaprow() \{
  state = (((state>>>6)&0x3) << 6)    // x1\('\) = x1
        | (((state>>>0)&0x3) << 4)    // x2\('\) = x4  *
        | (((state>>>2)&0x3) << 2)    // x3\('\) = x3
        | (((state>>>4)&0x3) << 0);   // x4\('\) = x2  *
\}
\end{program}

\noindent
Multiplication by matrix $M$ is done by the `mixcolumns' method:

\begin{program}
private void mixcolumns() \{
  int[] state_vector    = new int[4];  // 2 bits each entry
  int[] newstate_vector = new int[4];  // 2 bits each entry
  for (int i = 0; i < 4; i++, state >>>= 2)
    state_vector[i] = state & 0x3;
  // matrix multiplication on vectors of 4 groups of 2 bits
  //   1 2 0 0
  //   2 1 0 0
  //   0 0 1 2
  //   0 0 2 1
  newstate_vector[3] = state_vector[3]
                     ^ field_multiply(state_vector[2], 2);
  newstate_vector[2] = state_vector[2]
                     ^ field_multiply(state_vector[3], 2);
  newstate_vector[1] = state_vector[1]
                     ^ field_multiply(state_vector[0], 2);
  newstate_vector[0] = state_vector[0]
                     ^ field_multiply(state_vector[1], 2);
  for (int i = 3; i >= 0; i--)
    state = (state << 2) | newstate_vector[i];
\}
\end{program}

\noindent
The `inv\_mixcolumns' method multiplies by $M^{-1}$ instead, and is similar.
For completeness it is set out below:

\begin{program}
void inv_mixcolumns() \{
  int[] state_vector    = new int[4];
  int[] newstate_vector = new int[4];
  for (int i = 0; i < 4 ; i++, state >>>= 2)
    state_vector[i] = state & 0x3;
  // matrix multiplication on vectors of 4 groups of 2 bits
  //   3 1 0 0 
  //   1 3 0 0 
  //   0 0 3 1
  //   0 0 1 3
  newstate_vector[3] = field_multiply(state_vector[3], 3)
                     ^ field_multiply(state_vector[2], 1);
  newstate_vector[2] = field_multiply(state_vector[2], 3)
                     ^ field_multiply(state_vector[3], 1);
  newstate_vector[1] = field_multiply(state_vector[1], 3)
                     ^ field_multiply(state_vector[0], 1);
  newstate_vector[0] = field_multiply(state_vector[0], 3)
                     ^ field_multiply(state_vector[1], 1);
  for (int i = 3 ; i >= 0; i--)
    state = (state << 2) | newstate_vector[i];
\}
\end{program}

\noindent
The auxillary function `field\_multiply' does multiplication in the field
GF(4) of 2-bit numbers in the range 0-3. It regards $a$ and $b$  as vectors of 
bits, each of which is a coefficent of a power of $x$ in a polynomial. Then 
it multiplies the polynomials and returns the vector of bits which are
the coefficients of the powers of $x$ in the result polynomial. Recall
that $x^2=x+1$, i.e. $x^2+x+1=0$:

\begin{program}
static int field_multiply(int a, int b) \{

  int ret = 0;
  int[] xmultiples = new int[2];   // a, a*x, ...

  xmultiples[0] = a;
  for (int i = 1; i < 2; i++)      // a*x^n = x*(a*x^(n-1))
    xmultiples[i] = x_multiply(xmultiples[i - 1]);
  for (int i = 0; i < 2; i++) \{
    if ((b & 1) != 0)
      ret ^= xmultiples[i];        // b0*a+b1*a*x+... = b*a
    b >>>= 1;
  \}
  return ret;
\}
\end{program}

\noindent
The `x\_multiply' auxillary does the job of multiplication by 10
in the GF(4) field, which is multiplication by the polynomial $x$.
Recall that $x^2=x+1$:

\begin{program}
static int x_multiply(int a) \{
  switch (a) \{
    case 0: return 0;  // x*0 = 0
    case 1: return 2;  // x*1 = x
    case 2: return 3;  // x*x = x+1
    case 3: return 1;  // x*(x+1) = x^2+x = x+1+x = 1
  \}
  throw new NullPointerException();
\}
\end{program}

\noindent
The `substitute\_nibble' method applies the permutation $S$ to each
ordinate of the 4D state vector:

\begin{program}
void substitute_nibble() \{

  for (int i = 0; i < 4; i++) \{
      int t = s[state & 0x3]; // 2 bit
      state >>>= 2;
      state |= (t & 0x3) << 6;
  \}
\}
\end{program}

\noindent
The `s' array records the $S$ permutation on the elements of GF(4),
listing the image $S(i)$ in ${\rm s}[i]$ for each $i$ in the range 0-3.

The `inv\_substitute\_nibble' method applies the inverse
permutation $S^{-1}$:

\begin{program}
void inv_substitute_nibble() \{

  for (int i = 0; i < 4; i++) \{
      int t = inv_s[state & 0x3]; // 2 bit
      state >>>= 2;
      state |= (t & 0x3) << 6;
  \}
\}
\end{program}

\noindent
The `inv\_s' array records the $S^{-1}$ permutation on the elements of GF(4),
listing the image $S^{-1}(i)$ in ${\rm inv\_s}[i]$ for each $i$ in the
range 0-3.

To set up for encryption/decryption, one needs to first make an encryption
object, providing it with a randomly generated 8-bit key:

\begin{program}
static public void main(String [] args) \{

  byte key        = (byte)(256 * Math.random()); // 8-bit key
  simple_aes8 aes = new simple_aes8(key);
  byte in, out;
  \dots
  out = aes.ecb_encrypt(in);
  \dots
  in = aes.ecb_decrypt(out);
  \dots
\}
\end{program}

\noindent
The constructor for the encryption/decryption object makes the round keys
from the input key and embeds them in the object:

\begin{program}
simple_aes8(byte key) \{
  setkey(key);
\}
\end{program}

\noindent
The `setkey' method follows the prescription given 
for generating the round keys.

\begin{program}
void setkey(byte key) \{                           // 8 bit key

  int[] w   = new int[6];                         // 4 bits each
  roundkeys = new byte[3];                        // 3 instance variables

  w[0] = (key >>> 4) & 0xf;                       // upper 4 bits of key
  w[1] = key & 0xf;                               // lower 4 bits of key
  w[2] = w[0] ^ (2*4)                             // 4 bits
       ^ ((s[w[1] & 0x3] << 2) | s[w[1] >>> 2]);  // 4 bits

  w[3] = w[1] ^ w[2];                             // 4 bits
  w[4] = w[2] ^ (3 * 4)                           // 4 bits
       ^ ((s[w[3] & 0x3] << 2) | s[w[3] >>> 2]);  // 4 bits
  w[5] = w[3] ^ w[4];                             // 4 bits

  roundkeys[0] = (byte)(key & 0xff);              // 8 bits
  roundkeys[1] = (byte)((w[2] << 4) | w[3]);      // 8 bits
  roundkeys[2] = (byte)((w[4] << 4) | w[5]);      // 8 bits

  this.key = key;
\}
\end{program}

\noindent
As is to be seen above, the object has the following instance variables:

\begin{program}
  byte   key;          // input key, 8 bits
  byte[] roundkeys;    // 3 generated 'round keys', 8 bits each
  int state;           // working space, only 8 bits needed
\end{program}

\noindent
Additionally, the arrays `s' and `s\_inv'  for the $S$ permutation and
its inverse $S^{-1}$ are 
implemented as class constants.

\section{Addenda}

A belated look at the Wikipedia article on AES at {\small\url{http://en.wikipedia.org/wiki/Advanced_Encryption_Standard#High-level_description_of_the_algorithm}}
tells me that the permutation $S$ is supposed (for security) to have no
fixed points, so my 3-cycle on 0-3 in the 8-bit case is not recommended,
and best use the product of two transpositions, or a full 4-cycle.

The Wikipedia article also seems to say that the permutation $\pi$ of the
four ordinates is supposed to be a cycle, not a single transposition as
in the text above. In fact, the Wikipedia article seems to suggest that
the operation ought to be rather more complicated.

\section{Testing}

The following code can be used to establish if the implementation is
working right.

\begin{program}
static public void main(String [] args) \{
  byte key        = (byte)(256 * Math.random());
  int ntests      = 100;
  simple_aes8 aes = new simple_aes8(key);
  for (int i = 0; i < ntests; i++) \{
    byte plaintext1 = (byte)(256 * Math.random());
    byte ciphertext = aes.ecb_encrypt(plaintext1);
    byte plaintext2 = aes.ecb_decrypt(ciphertext);
    if (plaintext1 != plaintext2) \{
      System.err.print \dots
      System.exit(1);
    \}
  \}
\}
\end{program}

\noindent
For debugging, I recomemd printing out the (hex) value of the state
variable after each of the eight steps in the `ecb\_encrypt' and
`ecb\_decrypt' methods. They should read the same, in reverse order.
If there is a difference, that marks the step that goes wrong. Any problems
are likely due to Java code that pays insufficient attention to the
top fill after an implicit or explicit cast from a (negative-valued)
`byte' to an `int'. We want the top-fill to be zero always.

\section{12-bit code}

If you are going to adapt the above source to 12-bit blocks, you will
save yourself grief by changing all the `byte' declarations to `int' or
`short' at once (I prefer `short', as the compiler will then warn you
more often about what you might be doing wrong, which is better).  For
example, the testing routine could be:

\begin{program}
static public void main(String [] args) \{
  short key        = (short)(16 * 256 * Math.random());
  int ntests       = 100;
  simple_aes12 aes = new simple_aes12(key);
  for (int i = 0; i < ntests; i++) \{
    short plaintext1 = (short)(16 * 256 * Math.random());
    short ciphertext = aes.ecb_encrypt(plaintext1);
    short plaintext2 = aes.ecb_decrypt(ciphertext);
    if (plaintext1 != plaintext2) \{
      System.err.print \dots
      System.exit(1);
    \}
  \}
\}
\end{program}

\noindent
Where the 8-bit code has a for loop ranging over [0,2), the 12-bit
code needs a loop ranging over [0,3), as there are now three bits
in each ordinate of the 4D vectors. That applies to the
`field\_multiply' routine. 

Similarly, where the 8-bit code left- or right-shifts by 2 bits
at a time, the 12-bit code needs shifts of 3 bits at a time. That
applies to the `mixcolumns' and `inv\_mixcolumns' routines, and
`substitute\_nibble' and inv\_substitute\_nibble' routines, as well as
`swaprow'.

And where the 8-bit code masks with `\& 0x3', to get 2 bits out at a
time, the 12-bit code needs to mask with `\& 0x7', to get 3 bits out at
a time. That applies to `mixcolumns' and `inv\_mixcolumns', and
`substitute\_nibble' and inv\_substitute\_nibble' routines, and also
`swaprow'.

Similarly, the 12-bit `ecb\_encrypt' and `ecb\_decrypt' routines need to mask
by `\& 0xfff' for 12-bit input/output, not `\& 0xff' and 8-bit input/output.

The `x\_multiply' routine needs to be changed to reflect the
mutiplication-by-$x$ calculation in GF(8) instead of GF(4). Here is a
suitable code:

\begin{program}
static short x_multiply(short a) \{
  // 4 -> 3, i.e. x**2 -> x**3 = x+1, reflecting x**3 + x + 1 = 0
  return (short)((a < 4) ? 2 * a : (2 * a - 8) ^ 3);
\}
\end{program}

\noindent
Manage those changes correctly, and you should have a 12-bit block AES
cipher.

\section{Summary}
I have described AES encryption using 8-bit and/or 12-bit keys and blocks.
Sufficient Java source code has been set out that an implementation should
be a matter of copy, paste and edit.  To change the encryption other
than by varying the input key, modify the permutations $S$ and $\pi$,
and the (invertible) matrix $M$.

\newpage
\appendix
\newenvironment{smallprogram}{
\begin{scriptsize}\begin{quote}\begin{alltt}
}{
\end{alltt}\end{quote}\end{scriptsize}
}

\section{8-bit encryption/decryption Java source code}

\begin{smallprogram}
/*
 * 8-bit AES cipher.
 *
 * (C) Peter Breuer 2013 (ptb@inv.it.uc3m.es) for any parts I've written
 * myself, the whole of this source having been created by reverse
 * engineering some unattributed fragments of C for larger block AES which I
 * found publicly available on the web via Google with no licence or author
 * named inside (or anywhere around, under, over, etc) those sources.
 *
 * For the record those sources were
 *
 *   ecb_decrypt.c        806B
 *   ecb_encrypt.c        1089B
 *   simple_aes.c         2926B
 *   simple_aes_decr.c    3094B
 *
 * and if somebody can recognise and substantiate where those
 * ultimately come from, I'll be happy to acknowledge as appropriate.
 * The total of comments in those files is
 *
 *  ecb_decrypt.c // Encrypts input file to standard output.
 *  ecb_decrypt.c // Compile gcc ecb_encrypt.c simple_aes.o -o ecb
 *  ecb_decrypt.c // Usage:  ecb inputfile>outputfile
 *  ecb_decrypt.c // Encrypts input file to standard output.
 *  ecb_encrypt.c // Compile gcc ecb_encrypt.c simple_aes.o -o ecb
 *  ecb_encrypt.c // Usage:  ecb inputfile>outputfile
 *
 * I'm happy to place my code here under
 *
 *       * Gnu General Public Licence Version 2 (June 1991) *
 *
 * the required rubric for which is
 *
 *  This program is free software; you can redistribute it and/or modify
 *  it under the terms of the GNU General Public License as published by
 *  the Free Software Foundation; either version 2 of the License, or (at
 *  your option) any later version.
 *
 *  This program is distributed in the hope that it will be useful,
 *  but WITHOUT ANY WARRANTY; without even the implied warranty of
 *  MERCHANTABILITY or FITNESS FOR A PARTICULAR PURPOSE.  See the
 *  GNU General Public License for more details.
 *
 * To get a copy of the GPL2, search for "GPL", "GPL-2", "GPL2" on the
 * Internet, in particular at fsf.org.  Otherwise "write to the Free
 * Software Foundation, Inc., 51 Franklin Street, Fifth Floor, Boston,
 * MA 02110-1301 USA" for hardcopy.
 *
 * That licence means, paraphrasing, that you may use this source code
 * and change it and redistribute it in source and/or binary form, but you
 * must acknowledge where it comes from (i.e. include my name in the
 * history) and provide source on demand or by default to whoever you
 * distribute the binary to, and bind recipients of this or derived
 * source to the same or a compatible licence. That means that they are
 * free to change it, have to bind recipients of their binary or source
 * to the same or a compatible licence, etc. The upshot is that people
 * who receive the compiled code always also get the right to change
 * the source to suit themselves, and that right is inherited through
 * any number of derivations and hand-offs. 
 * 
 * Not upholding your end of the licence terms means that the licence
 * is automatically revoked, and then your use is governed by copyright
 * law (which essentally means no use without permission, except for fair
 * use exceptions as determined by copyright law).
 */

/*
 * This file contains the Java simple_aes8 class, with its methods
 *
 *         simple_aes8(byte)    // makes a de/encryption device with  key 
 *    byte ecb_encrypt(byte)    // encrypt one byte using key
 *    byte ecb_decrypt(byte)    // decrypt one byte using key
 *         setkey(byte)         // reset key
 *         setverbose()         // make debugging noise
 *         setquiet()           // make no debugging noise
 *
 * note: 8-bit key, 8-bit block.
 */
public class simple_aes8 \{

    // 0->3->2->0; 1->1;
    private static int[] sbox = \{           // permutation used in encryption
        3,1,0,2,
    \};
    // 0->2->3->0; 1-1
    private static int[] invsbox = \{        // inverse permutation
        2,1,3,0
    \};

    private byte key;                       // 8-bit key
    private byte[] roundkeys = null;        // derived keys

    private int state;                      // modified by en/decryption
    private boolean verbose = false;

    private int interstate[] = new int[9];  // debugging encryption process
    private int inverstate[] = new int[9];  // debugging decryption process

    /*
     * Multiplication in GF(4).  Field elements are integers in the range
     * 0...3, which we think of as degree < 2 polynomials over GF(2).
     *  a1 a0 * b1 b0 = 
     *  .. (b1b0+b1b1+b0b1)(a0b0+b1b1)
     * (X**2 = X+1) since X**2+X+1 is irreducible mod 2.
     */
    private static int field_multiply(int a, int b) \{

        int ret = 0;
        int[] xmultiples = new int[2]; // a, a*x

        xmultiples[0] = a;

        for (int i = 1; i < 2; i++)  // a*x^1 = x*(a*x^0)
             xmultiples[i] = x_multiply(xmultiples[i - 1]);

        for (int i = 0; i < 2; i++) \{
             if ((b & 1) != 0)
                 ret ^= xmultiples[i]; // b0*a+b1*a*x = b*a
             b >>>= 1;
        \}
        return ret;
    \}

    /*
     * Multiplication by the element x (= 1*x+0*1 = "10") of the field
     */
    private static int x_multiply(int a) \{

        // 4 -> 3, i.e. x**2 -> x+1, so x**2 = x+1
        switch (a) \{
           case 0:
             return 0;  // x*0 = 0
           case 1:      // x*1 = x
             return 2;
           case 2:
             return 3;  // x*x = x+1
           case 3:
             return 1;  // x*(x+1) = x^2+x = 1
        \}
        throw new NullPointerException();
    \}

    /*
     * Make and install the derived keys from the original key
     */
    void setkey(byte key) \{

        if (this.key == key && roundkeys != null)
            return;

        int[] w = new int[6];       // 4 bits each

        roundkeys = new byte[3];

        w[0] = (key >>> 4) & 0xf;   // upper 4 bits of key
        w[1] = key & 0xf;           // lower 4 bits of key
        w[2] = w[0] ^ (2*4)         // 4 bits
             ^ ((sbox[w[1] & 0x3] << 2) | sbox[w[1] >>> 2]); // 4 bits

        w[3] = w[1] ^ w[2];         // 4 bits
        w[4] = w[2] ^ (3 * 4)       // 4 bits
             ^ ((sbox[w[3] & 0x3] << 2) | sbox[w[3] >>> 2]); // 4 bits
        w[5] = w[3] ^ w[4];         // 4 bits

        roundkeys[0] = (byte)(key & 0xff);          // 8 bits
        roundkeys[1] = (byte)((w[2] << 4) | w[3]);  // 8 bits
        roundkeys[2] = (byte)((w[4] << 4) | w[5]);  // 8 bits

        // those roundkeys seem to have independent 4 bit components

        pdebug("round keys: \%04o{\textbackslash}t\%04o{\textbackslash}t\%04o{\textbackslash}n",
            roundkeys[0], roundkeys[1], roundkeys[2]);

        this.key = key;
    \}

    /*
     * Apply a subsitution in each of 4 groups of 2 bits each ("nibbles")
     * to the state.
     */
    private void substitute_nibble() \{

        // apply sbox permutation to each nibble

        int[] state_vector = new int[4]; // each is 2 bit!
        int newstate = 0;

        for (int i = 0; i < 4; i++) \{
            state_vector[i] = sbox[state & 0x3]; // 2 bit
            state >>>= 2;
        \}
        for (int i = 3; i >= 0; i--)
            newstate = (newstate << 2) | state_vector[i];
        state = newstate;
    \}

    /*
     * Apply inverse subsitution in each of 4 groups of 2 bits each ("nibbles")
     * to the state.
     */
    private void inv_substitute_nibble() \{

        // apply inverse sbox permutation to each nibble

        int[] state_vector = new int[4];
        int newstate = 0;

        for (int i = 0; i < 4; i++) \{
            state_vector[i] = invsbox[state & 0x3];  // 2 bit
            state >>= 2;
        \}
        for (int i = 3; i >= 0; i--)
            newstate = (newstate << 2) | state_vector[i];
        state = newstate;
    \}

    /*
     * Permute the nibbles. (x0,x1,x2,x3) -> (x2,x3,x0,x3)
     * to the state.
     */
    private void swaprow() \{

        int[] state_vector = new int[4];
        int newstate = 0;

        for (int i = 0; i < 4; i++) \{
            state_vector[i] = state & 0x3;
            state >>= 2;
        \}
        // swaps 0<->2 groups of 2 bits
        // 1 0 0 0
        // 0 0 0 1
        // 0 0 1 0
        // 0 1 0 0
        newstate = (state_vector[3] << 6)
                 | (state_vector[0] << 4)
                 | (state_vector[1] << 2)
                 | (state_vector[2] << 0);
        state = newstate;
    \}

    /*
     * Apply a linear transform to the state as a vector of 4 nibbles
     */
    private void mixcolumns() \{

        int[] state_vector = new int[4];  // 2 bits each
        int newstate = 0;
        int oldstate = state;
        int[] newstate_vector = new int[4];

        for (int i = 0; i < 4; i++) \{
            state_vector[i] = state & 0x3;  // 2 bits
            state >>>= 2;
        \}

        // matrix multiplication on groups of 2 bits
        //   1 2 0 0
        //   2 1 0 0
        //   0 0 1 2
        //   0 0 2 1
        newstate_vector[3] = state_vector[3]
                           ^ field_multiply(state_vector[2], 2);
        newstate_vector[2] = state_vector[2]
                           ^ field_multiply(state_vector[3], 2);
        newstate_vector[1] = state_vector[1]
                           ^ field_multiply(state_vector[0], 2);
        newstate_vector[0] = state_vector[0]
                           ^ field_multiply(state_vector[1], 2);

        for (int i = 3; i >= 0; i--)
            newstate = (newstate << 2) | newstate_vector[i];
        state = newstate;

    \}

    /*
     * Apply inverse linear transform to the state as a vector of 4 nibbles
     */
    private void inv_mixcolumns() \{

        int[] state_vector = new int[4];
        int   newstate = 0;
        int[] newstate_vector = new int[4];

        for (int i = 0; i < 4 ; i++) \{
            state_vector[i] = state & 0x3;
            state >>>= 2;
        \}
        // matrix multiplication on groups of 2 bits
        //   3 1 0 0 
        //   1 3 0 0 
        //   0 0 3 1
        //   0 0 1 3
        newstate_vector[3] = field_multiply(state_vector[3], 3)
                           ^ field_multiply(state_vector[2], 1);
        newstate_vector[2] = field_multiply(state_vector[2], 3)
                           ^ field_multiply(state_vector[3], 1);
        newstate_vector[1] = field_multiply(state_vector[1], 3)
                           ^ field_multiply(state_vector[0], 1);
        newstate_vector[0] = field_multiply(state_vector[0], 3)
                           ^ field_multiply(state_vector[1], 1);

        for (int i = 3 ; i >= 0; i--)
            newstate = (newstate << 2) | newstate_vector[i];
        state = newstate;
    \}

    /*
     * The debug generic printout routine. Only makes noise if verbose
     * set.
     */
    private void pdebug(String format, int ... args) \{

        if (!verbose)
            return;

        switch (args.length) \{
          case 0:
            System.out.printf(format);
            break;
          case 1:
            System.out.printf(format, args[0]);
            break;
          case 2:
            System.out.printf(format, args[0], args[1]);
            break;
          case 3:
            System.out.printf(format, args[0], args[1], args[2]);
            break;
          case 4:
            System.out.printf(format, args[0], args[1], args[2], args[3]);
            break;
        \}
    \}

    /*
     * encryption method applied to state
     */
    private void encrypt() \{

        // 0
        pdebug("E state: \%04o{\textbackslash}n", state);
        interstate[0] = state;

        // 1
        state ^= roundkeys[0];
        pdebug("E Add round key: \%04o{\textbackslash}n", state);
        interstate[1] = state;

        // 2
        substitute_nibble();     // code groups of 4 bits
        pdebug("E Substitute: \%04o{\textbackslash}n", state);
        interstate[2] = state;

        // 3
        swaprow();              // swap 0,2 groups
        pdebug("E Swap rows: \%04o{\textbackslash}n", state);
        interstate[3] = state;

        // 4
        mixcolumns();            // matrix multiply, preserves groups of 8 bits
        pdebug("E Mix Columns: \%04o{\textbackslash}n", state);
        interstate[4] = state;

        // 5
        state ^= roundkeys[1];   // hurr .. add a constant
        pdebug("E Add round key: \%04o{\textbackslash}n", state);
        interstate[5] = state;

        // 6
        substitute_nibble();     // SECOND coding!
        pdebug("E Substitute: \%04o{\textbackslash}n", state);
        interstate[6] = state;

        // 7
        swaprow();              // swap 0,2 groups
        pdebug("E Swap rows: \%04o{\textbackslash}n", state);
        interstate[7] = state;

        // 8
        state ^= roundkeys[2];   // .. add another constant
        pdebug("E Add round key: \%04o{\textbackslash}n", state);
        interstate[8] = state;
    \}  

    /*
     * decryption method applied to state
     */
    private void decrypt() \{

        // 0
        pdebug("D state: \%04o{\textbackslash}n", state);
        inverstate[0] = state;

        // 1
        state ^= roundkeys[2];
        pdebug("D Add round key: \%04o{\textbackslash}n", state);
        inverstate[1] = state;

        // 2
        swaprow();
        pdebug("D Swap rows: \%04o{\textbackslash}n", state);
        inverstate[2] = state;

        // 3
        inv_substitute_nibble();
        pdebug("D Substitute: \%04o{\textbackslash}n", state);
        inverstate[3] = state;

        // 4
        state ^= roundkeys[1];   // hurr .. add a constant
        pdebug("D Add round key: \%04o{\textbackslash}n", state);
        inverstate[4] = state;

        // 5
        inv_mixcolumns();
        pdebug("D Mix Columns: \%04o{\textbackslash}n", state);
        inverstate[5] = state;

        // 6
        swaprow();
        pdebug("D Swap rows: \%04o{\textbackslash}n", state);
        inverstate[6] = state;

        // 7
        inv_substitute_nibble();
        pdebug("D Substitute: \%04o{\textbackslash}n", state);
        inverstate[7] = state;

        // 8
        state ^= roundkeys[0];
        pdebug("D Add round key: \%04o{\textbackslash}n", state);
        inverstate[8] = state;

    \} 

    /*
     * set in order to make more noise
     */
    public void setverbose() \{
        verbose = true;
    \}

    /*
     * set in order to make less noise
     */
    public void setquiet() \{
        verbose = false;
    \}

    /*
     * encryption method applied to a 8-bit plaintext
     */
    public byte ecb_encrypt(byte input_block) \{

        state = input_block & 0xff;
        encrypt();
        return (byte)state;
    \}

    /*
     * decryption method applied to a 8-bit ciphertext
     */
    public byte ecb_decrypt(byte cipherblock) \{

        state = cipherblock & 0xff;
        decrypt();
        return (byte)state;
    \}

    /*
     * constructor for a cipher object from a 8-bit key
     */
    public simple_aes8(byte key) \{
        setkey(key);
    \}

    /*
     * encrypt and decrypt random 8-bit text 100 times
     */
    static public void main(String [] args) \{

        // 8-bit key
        byte key    = (byte)(256 * Math.random());
        int ntests  = 1000;
        int errs    = 0;
        simple_aes8 aes 
                    = new simple_aes8(key);

        System.out.println("Testing 8-bit encryption/decryption:");
        for (int i = 0; i < ntests; i++) \{

            // 8-bit text
            byte text1       = (byte)(256 * Math.random());
            byte ciphertext1 = (byte)(aes.ecb_encrypt(text1) & 0xff);
            byte text2       = aes.ecb_decrypt(ciphertext1);
            if (text1 != text2) \{

               System.out.printf("{\textbackslash}nmistake with key \%04o{\textbackslash}n", key);
               System.out.printf("in: \%d, out:\%d{\textbackslash}n",
                   text1 & 255, text2 & 255);
               errs++;
            \}
        \}
        System.out.println("{\textbackslash}r" + errs + "/" + ntests + " errors"); 
        if (errs > 0)
               System.exit(1);

    \}
\}   // end of class 
\end{smallprogram}

\newpage

\section{12-bit encryption/decryption Java source code}

\begin{smallprogram}
/*
 * 12-bit AES cipher.
 *
 * (C) Peter Breuer 2013 (ptb@inv.it.uc3m.es) for any parts I've written
 * myself, the whole of this source having been created by reverse
 * engineering unattributed fragments publicly available on
 * the web with no licence or author named inside them. I'm happy to place
 * what's here under
 *
 *       * Gnu General Public Licence Version 2 (June 1991) *
 *
 * the required rubric for which is
 *
 *  This program is free software; you can redistribute it and/or modify
 *  it under the terms of the GNU General Public License as published by
 *  the Free Software Foundation; either version 2 of the License, or (at
 *  your option) any later version.
 *
 *  This program is distributed in the hope that it will be useful,
 *  but WITHOUT ANY WARRANTY; without even the implied warranty of
 *  MERCHANTABILITY or FITNESS FOR A PARTICULAR PURPOSE.  See the
 *  GNU General Public License for more details.
 *
 * To get a copy of the GPL2, search for "GPL", "GPL-2", "GPL2" on the
 * Internet, in particular at fsf.org.  Otherwise "write to the Free
 * Software Foundation, Inc., 51 Franklin Street, Fifth Floor, Boston,
 * MA 02110-1301 USA" for hardcopy.
 *
 * That licence means, paraphrasing, that you may use this source code
 * and change it and redistribute it in source and/or binary form, but you
 * must acknowledge where it comes from (i.e. include my name in the
 * history) and provide source on demand or by default to whoever you
 * distribute the binary to, and bind recipients of this or derived
 * source to the same or a compatible licence. That means that they are
 * free to change it, have to bind recipients of their binary or source
 * to the same or a compatible licence, etc. The upshot is that people
 * who receive the compiled code always also get the right to change
 * the source to suit themselves, and that right is inherited through
 * any number of derivations and hand-offs. 
 * 
 * Not upholding your end of the licence terms means that the licence
 * is automatically revoked, and then you are goverened by copyright
 * law in using this source (which essentally means no use without
 * permission, except for fair use exceptions as determined by copyright
 * law).
 */

/*
 * This file contains the Java simple_aes12 class, with its methods
 *
 *         simple_aes12(short)   // makes a de/encryption device with key 
 *   short ecb_encrypt(short)    // encrypt byte using key
 *   short ecb_decrypt(short)    // decrypt byte using key
 *         setkey(short)         // reset key
 *         setverbose()          // make debugging noise
 *         setquiet()            // make no debugging noise
 *
 * note: 12-bit key, 12-bit block.
 */
class simple_aes12 \{

    // 0->2->6->0; 1->4->7->5->3->1;
    private static int[] sbox = \{           // permutation used in encryption
        2,4,6,1,
        7,3,0,5,
    \};
    // 0->6->2->0; 1->3->5->7->4->1
    private static int[] invsbox = \{        // inverse permutation
        6,3,0,5,
        1,7,2,4,
    \};

    // 12-bit key
    private short key;                      // 12-bit key
    private short[] roundkeys = null;       // derived keys

    private int state;
    private boolean verbose = false;

    /*
     * Multiplication in GF(8).  Field elements are integers in the range
     * 0...7, which we think of as degree < 3 polynomials over GF(2).
     *  a3 a2 a1 a0 * b3 b2 b1 b0 = 
     *  .. (b2b0+b1b1+b0b2+b3b3)(b0b1+b1b0+b3b3+b2b3)(b0b0+b3b1+b2b2+b1b3)
     * (X**3 = X+1) since X**3+X+1 is irreducible mod 2.
     */
    private static int field_multiply(int a, int b) \{

        int ret = 0;
        int[] xmultiples = new int[4];

        xmultiples[0] = a;

        for (int i = 1; i < 4; i++)
             xmultiples[i] = x_multiply(xmultiples[i - 1]);

        for (int i = 0; i < 4; i++) \{
             if ((b & 1) != 0)
                 ret ^= xmultiples[i];
             b >>>= 1;
        \}
        return ret;
    \}

    /*
     * Multiplication by the element x (= 0*x^2+1*x+0*1 = "010") of the field
     */
    private static int x_multiply(int a) \{

        // 8 -> 3, i.e. x**3 -> x+1, so x**3 = x+1
        return (a < 4) ? 2 * a : (2 * a - 8) ^ 3;
    \}

    /*
     * Make and install the derived keys from the original 12-bit key
     */
    void setkey(short key) \{

        if (this.key == key && roundkeys != null)
            return;

        int[] w = new int[6];       // 6 bits each

        roundkeys = new short[3];

        w[0] = (key >>> 6) & 0x3f;  // upper 6 bits of key
        w[1] = key & 0x3f;          // lower 6 bits of key
        w[2] = w[0] ^ (4*8)         // 6 bits
             ^ ((sbox[w[1] & 0x7] << 3) | sbox[w[1] >>> 3]); // 6 bits
        w[3] = w[1] ^ w[2];         // 6 bits
        w[4] = w[2] ^ (3 * 8)       // 6 bits
             ^ ((sbox[w[3] & 0x7] << 3) | sbox[w[3] >>> 3]); // 6 bits
        w[5] = w[3] ^ w[4];         // 6 bits

        roundkeys[0] = key;         // 12 bits
        roundkeys[1] = (short)((w[2] << 6) | w[3]);  // 12 bits
        roundkeys[2] = (short)((w[4] << 6) | w[5]);  // 12 bits

        // those roundkeys seem to have independent 4 bit components

        pdebug("round keys: %04o{\textbackslash}t%04o{\textbackslash}t%04o{\textbackslash}n",
            roundkeys[0], roundkeys[1], roundkeys[2]);

        this.key = key;
    \}

    /*
     * Apply a subsitution in each of 4 groups of 3 bits each ("nibbles")
     * to the state.
     */
    private void substitute_nibble() \{

        int[] state_vector = new int[4]; // each is 3 bit!
        int newstate = 0;

        for (int i = 0; i < 4; i++) \{
            state_vector[i] = sbox[state & 0x7]; // 3 bit
            state >>>= 3;
        \}
        for (int i = 3; i >= 0; i--)
            newstate = (newstate << 3) | state_vector[i];
        state = newstate;
    \}

    /*
     * Apply inverse subsitution in each of 4 groups of 3 bits each ("nibbles")
     * to the state.
     */
    private void inv_substitute_nibble() \{

        int[] state_vector = new int[4];
        int newstate = 0;

        for (int i = 0; i < 4; i++) \{
            state_vector[i] = invsbox[state & 0x7];
            state >>>= 3;
        \}
        for (int i = 3; i >= 0; i--)
            newstate = (newstate << 3) | state_vector[i];
        state = newstate;
    \}

    /*
     * Permute the nibbles. (x0,x1,x2,x3) -> (x2,x3,x0,x3)
     * to the state.
     */
    private void swaprow() \{

        int[] state_vector = new int[4];
        int newstate = 0;

        for (int i = 0; i < 4; i++) \{
            state_vector[i] = state & 0x7;
            state >>>= 3;
        \}
        // swaps 0<->2 groups of 3 bits
        // 1 0 0 0
        // 0 0 0 1
        // 0 0 1 0
        // 0 1 0 0
        newstate = (state_vector[3] << 9)
                 | (state_vector[0] << 6)
                 | (state_vector[1] << 3)
                 | (state_vector[2] << 0);
        state = newstate;
    \}

    /*
     * Apply a linear transform to the state as a vector of 4 nibbles
     */
    private void mixcolumns() \{

        int[] state_vector = new int[4];
        int newstate = 0;
        int oldstate = state;
        int[] newstate_vector = new int[4];

        for (int i = 0; i < 4; i++) \{
            state_vector[i] = state & 0x7;
            state >>>= 3;
        \}

        // matrix multiplication on groups of 3 bits
        //   1 4 0 0
        //   4 1 0 0
        //   0 0 1 4
        //   0 0 4 1
        newstate_vector[3] = state_vector[3]
                           ^ field_multiply(state_vector[2], 4);
        newstate_vector[2] = state_vector[2]
                           ^ field_multiply(state_vector[3], 4);
        newstate_vector[1] = state_vector[1]
                           ^ field_multiply(state_vector[0], 4);
        newstate_vector[0] = state_vector[0]
                           ^ field_multiply(state_vector[1], 4);

        for (int i = 3; i >= 0; i--)
            newstate = (newstate << 3) | newstate_vector[i];
        state = newstate;

    \}

    /*
     * Apply inverse linear transform to the state as a vector of 4 nibbles
     */
    private void inv_mixcolumns() \{

        int[] state_vector = new int[4];
        int newstate = 0;
        int[] newstate_vector = new int[4];

        for (int i = 0; i < 4 ; i++) \{
            state_vector[i] = state & 0x7;
            state >>>= 3;
        \}
        // matrix multiplication on groups of 3 bits
        //   4 6 0 0
        //   6 4 0 0
        //   0 0 4 6
        //   0 0 6 4
        newstate_vector[3] = field_multiply(state_vector[3], 4)
                           ^ field_multiply(state_vector[2], 6);
        newstate_vector[2] = field_multiply(state_vector[2], 4)
                           ^ field_multiply(state_vector[3], 6);
        newstate_vector[1] = field_multiply(state_vector[1], 4)
                           ^ field_multiply(state_vector[0], 6);
        newstate_vector[0] = field_multiply(state_vector[0], 4)
                           ^ field_multiply(state_vector[1], 6);

        for (int i = 3 ; i >= 0; i--)
            newstate = (newstate << 3) | newstate_vector[i];
        state = newstate;
    \}

    /*
     * The debug generic printout routine. Only makes noise if verbose
     * set.
     */
    private void pdebug(String format, int ... args) \{

        if (!verbose)
            return;

        switch (args.length) \{
          case 0:
            System.out.printf(format);
            break;
          case 1:
            System.out.printf(format, args[0]);
            break;
          case 2:
            System.out.printf(format, args[0], args[1]);
            break;
          case 3:
            System.out.printf(format, args[0], args[1], args[2]);
            break;
          case 4:
            System.out.printf(format, args[0], args[1], args[2], args[3]);
            break;
        \}
    \}

    /*
     * encryption method applied to state
     */
    private void encrypt() \{

        // 0
        pdebug("E state: %04o{\textbackslash}n", state);

        // 1
        state ^= roundkeys[0];
        pdebug("E Add round key: %04o{\textbackslash}n", state);

        // 2
        substitute_nibble();  // code groups of 3 bits
        pdebug("E Substitute: %04o{\textbackslash}n", state);

        // 3
        swaprow();           // swap 0,2 groups
        pdebug("E Swap rows: %04o{\textbackslash}n", state);

        // 4
        mixcolumns();         // matrix multiply, preserves groups of 8 bits
        pdebug("E Mix Columns: %04o{\textbackslash}n", state);

        // 5
        state ^= roundkeys[1];   // hurr .. add a constant
        pdebug("E Add round key: %04o{\textbackslash}n", state);

        // 6
        substitute_nibble();  // SECOND coding!
        pdebug("E Substitute: %04o{\textbackslash}n", state);

        // 7
        swaprow();           // swap 0,2 groups
        pdebug("E Swap rows: %04o{\textbackslash}n", state);

        // 8
        state ^= roundkeys[2];   // .. add another constant
        pdebug("E Add round key: %04o{\textbackslash}n", state);
    \}  

    /*
     * decryption method applied to state
     */
    private void decrypt() \{

        // 0
        pdebug("D state: %04o{\textbackslash}n", state);

        // 1
        state ^= roundkeys[2];
        pdebug("D Add round key: %04o{\textbackslash}n", state);

        // 2
        swaprow();
        pdebug("D Swap rows: %04o{\textbackslash}n", state);

        // 3
        inv_substitute_nibble();
        pdebug("D Substitute: %04o{\textbackslash}n", state);

        // 4
        state ^= roundkeys[1];   // hurr .. add a constant
        pdebug("D Add round key: %04o{\textbackslash}n", state);

        // 5
        inv_mixcolumns();
        pdebug("D Mix Columns: %04o{\textbackslash}n", state);

        // 6
        swaprow();
        pdebug("D Swap rows: %04o{\textbackslash}n", state);

        // 7
        inv_substitute_nibble();
        pdebug("D Substitute: %04o{\textbackslash}n", state);

        // 8
        state ^= roundkeys[0];
        pdebug("D Add round key: %04o{\textbackslash}n", state);

    \} 

    /*
     * set in order to make more noise
     */
    public void setverbose() \{
        verbose = true;
    \}

    /*
     * set in order to make less noise
     */
    public void setquiet() \{
        verbose = false;
    \}

    /*
     * encryption method applied to a 12-bit plaintext
     */
    public short ecb_encrypt(short input_block) \{

        state = input_block & 0xfff;
        encrypt();
        return (short)(state & 0xfff);
    \}

    /*
     * decryption method applied to a 12-bit ciphertext
     */
    public short ecb_decrypt(short cipherblock) \{

        state = cipherblock & 0xfff;
        decrypt();
        return (short)(state & 0xfff);
    \}

    /*
     * constructor for a cipher object from a 12-bit key
     */
    public simple_aes12(short key) \{
        setkey(key);
    \}

    /*
     * encrypt and decrypt random 12-bit text 1000 times
     */
    static public void main(String [] args) \{

        // 12-bit key
        short key   = (short)(64 * 64 * Math.random());
        int ntests  = 1000;
        simple_aes12 aes 
                    = new simple_aes12(key);
        int errs = 0;

        //aes.setverbose();

        System.out.println("Testing 12-bit encryption/decryption:");
        for (int i = 0; i < ntests; i++) \{

            // 12-bit text
            short text1       = (short)(64 * 64 * Math.random());
            short ciphertext1 = aes.ecb_encrypt(text1);
            short text2       = aes.ecb_decrypt(ciphertext1);

            if (text1 != text2) \{

               System.out.printf("mistake with key %04o{\textbackslash}n", key);
               System.out.printf("in: %d, out:%d{\textbackslash}n",
                   text1 & 0xfff, text2 & 0xfff);
               errs++;
            \}
        \}
        System.out.println("" + errs + "/" + ntests + " errors"); 
        if (errs > 0)
               System.exit(1);
    \}
\} // end of class
\end{smallprogram}

\end{document}